# Anticipatory Structure in the Propagation of Signal

M.R. Sayeh and R.E. Auxier
Southern Illinois University
Carbondale, Illinois 62901

*Abstract*

We here report the development of a structure that shows the proteresis phenomenon in more general setting and set out its philosophical implications. In this case, the questions relate to how we are to interpret what will happen in the future, and the "procollection" (the counterpart of "recollection") of not-yet-experienced phenomena that, when expressed, *will be* whatever has built up in fully determinate form *already*, ahead of the event. If such a process really exists, as our evidence confirms, not just as phenomenon but as a fact, then a gap exists between the actualized form of the future and its concrete expression when the event does "happen." Such a fact, as hard to imagine as it is, may be intelligible, even interpretable and susceptible to mathematical and/or logical modeling. We build upon neglected theories and formulae that present time in a way that makes our results interpretable. A proteretic device is here described which shifts the input signal (event) to the future – it is an "anticipatory structure." The proteretic characteristic of neurons should also be capable of demonstration; and its neuronal behavior is possibly the reason for the fast perception/thought processes in spite of slow behaving neurons. That capacity may also account for why it is possible for animals (including humans) to interact with the environment by slightly "seeing" (in the sense of perceiving and/or sensing) the future. Exploiting this new proteretic technology, faster computers and more efficient cellphones, among other things, will be designed and built.

## Introduction

We will explore a phenomenon we call "proteresis," which is a futural analogue of hysteresis. The hysteresis phenomenon relates to delay, what happens in the past, with memory, and other physiological and more general biological processes, as past experience is stored without being released in action. But hysteresis is not limited to living systems. A hysteretic "bulge" is evident, for example, in the difference between what one sees and what one hears when a baseball is hit and one is standing a hundred meters away. It is the same "event," but its expression *in* sound waves and *as* light creates a difference that requires the sound one hears to "catch up" to what one sees as the ball approaches the place one is standing. It is not simply a function of the observer; the light is actually travelling much faster than the sound. They "come apart," so to speak, the sound "builds up" behind the light, but then "rejoins" as the event ends.[1]

What if the phenomenon of hysteresis exists in reverse? Is there also *pro*teresis in non-living systems? We have found that it does happen. We here report the development of a structure that shows it. In this case, the questions relate to how we are to interpret the "negative delay," what will happen in future, and the "procollection" of not yet experienced[2] phenomena that, when expressed, *will be* whatever has built up, in fully determinate form, *already*. If such a process really exists, as our evidence confirms, not just as phenomenon but as fact, the gap closes between the actualized form of the future and its expression when it does "happen." Such a fact may be intelligible, even interpretable and susceptible to mathematical and/or logical modeling. A proteretic device is here described which shifts the input signal (event) to the future – it is an "anticipatory structure."

Mathematical study of the hysteresis phenomenon is very difficult; and a systematic and rigorous study has been initiated few decades ago by Russian mathematicians, M. A. Krasnoselskii and A. V. Pokrovskii,[1] see also [2,3,4,5]. We may expect still greater difficulties with proteresis. The challenge has obliged us to comb through the history of mathematics for formalizations from the time before the standard model of gravitational cosmology (i.e., general relativity and everything associated with the study of it) came to be so dominant as to suppress interest in alternative formalizations. We have settled on a relatively complete set of formalizations that will be referenced here but explained in a subsequent paper. The aim at present is to describe the machine itself and offer some philosophical ideas for interpreting what it does.

The proteresis "phenomenon" (if it can be called that, since what we seek is its existence in fact, not just as appearance) has been mentioned only in pharmaceutical related fields [11] and recently in ferroelectric and ferromagnetic materials [12,13]. Proteretic bi-stable (binary) devices have been implemented electronically [10] as well as optically [7]. In this paper, we carry the proteretic idea further to multinary systems and set out its philosophical implication.

## *Mathematical System Perspective*

In this section we discuss a structure which gives rise to the proteretic behavior as applicable in physical devices. As shown in Figure 1, this proteretic device consists of two hysteretic bi-stable devices and a feedforward connection – the need for a feedforward connection for the anticipatory systems has been mentioned by Robert Rosen [6].[3] The A device has to be an inverting one; and the B device takes either inverting or noninverting (shown) type. Figure 2 shows the proposed proteretic device symbol.

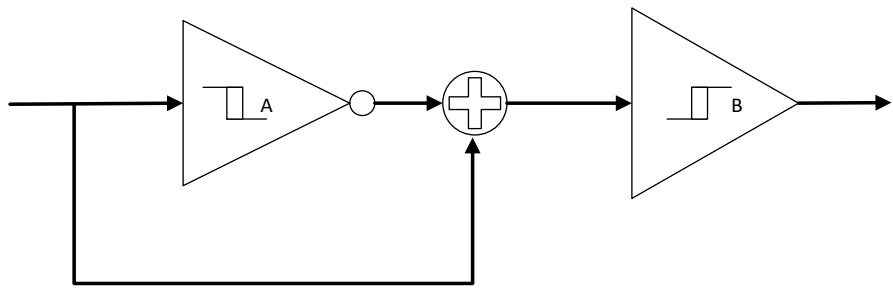

Figure 1. Proteretic device structure.

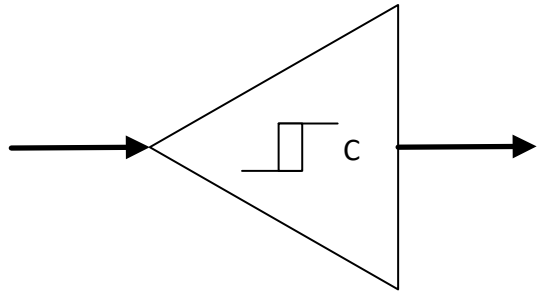

Figure 2. Proteretic device symbol.

For device A (B) in Figure 1, assuming the on-threshold value and the off-threshold value to be $a_A$ ($b_B$) and $b_A$ ($a_B$), respectively, the on-threshold value and the off-threshold value for device C will be

$$a_C = a \text{ and } b_C = b, \text{respectively,}$$

given $a_A = a$, $b_A = b$, $a_B = b$, and $b_B = 1 + a$. The aforementioned analysis based on the unitless binary values of one and zero [7].

This idea of binary proteretic device is extended to a multinary (M-ary) proteretic system as shown in Figure 3 and symbolized in Figure 4. .

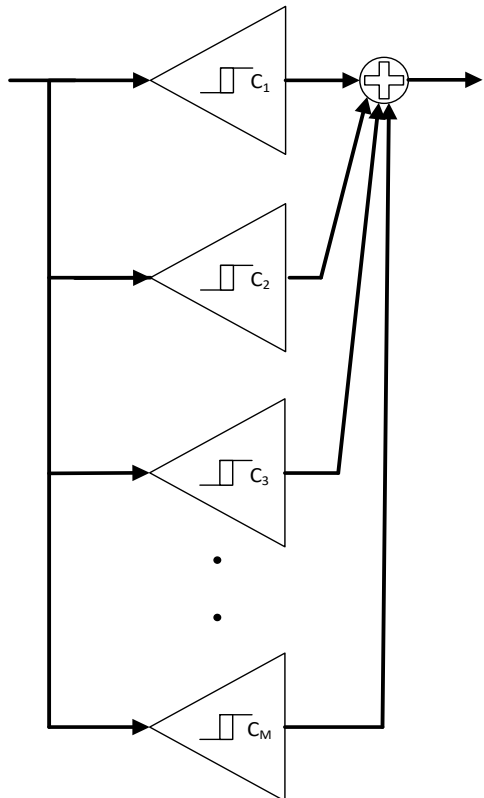

Figure 3. Multinary (M-ary) proteretic system.

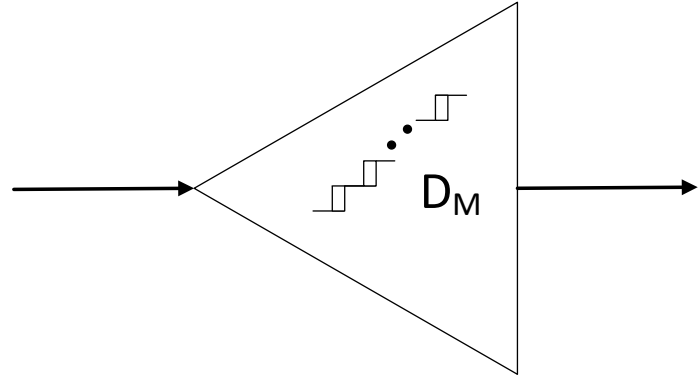

Figure 4. Multinary proteretic system symbol.

Using a twelve-nary (dodecanary) system ($M = 12$), a simulation is performed (via Simulink) with an almost random signal input. The transfer function (input vs. output) is shown in Figure 5 - note the direction of the arrows in the opposing direction of the hysteretic system. In consequence, the output shows a shift into the future, see Figure 6. The blue (red) signal is the input (output). For this example, $a_{D_m} = m - 0.8$ and $b_{D_m} = m - 0.2$, where $m$ = 1, 2, 3, …, 12. As shown, the output is a quantized version of the input, shifted to the future by a negative delay t∆.[4] (Perhaps "negative delay" should be written as a "prolay." It is difficult to choose adequate language.)

Assuming $a_{D_m} + b_{D_m} = m$, the negative delay, t∆, is given as

$$t\Delta = \frac{b_{D_m} - a_{D_m}}{2|R|},$$

where $R$ represents the rate of change for the input signal at "present" time (whenever we may designate between time $t_n$ in the future, and time $t_1$ which is the present). Figure 7 depicts three noisy input pulses with different rates and their output responses, showing the dependency of t∆ to the input rate as well as the noise immunity. The hysteretic devices used in this work are time independent, meaning their transfer functions are time independent of one another. Of course, this scheme can be extended to time dependent or rate dependent cases.

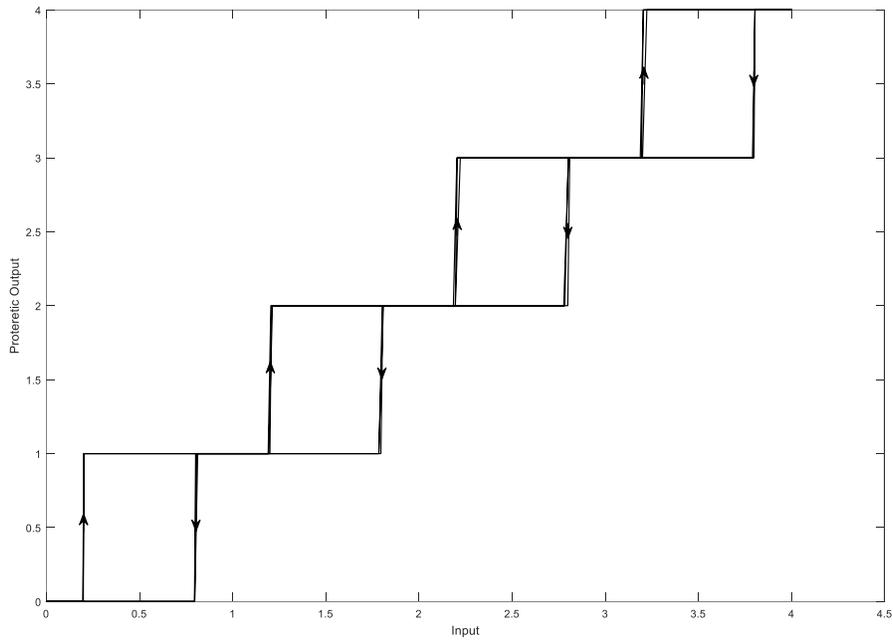

Figure 5. Proteretic transfer function.

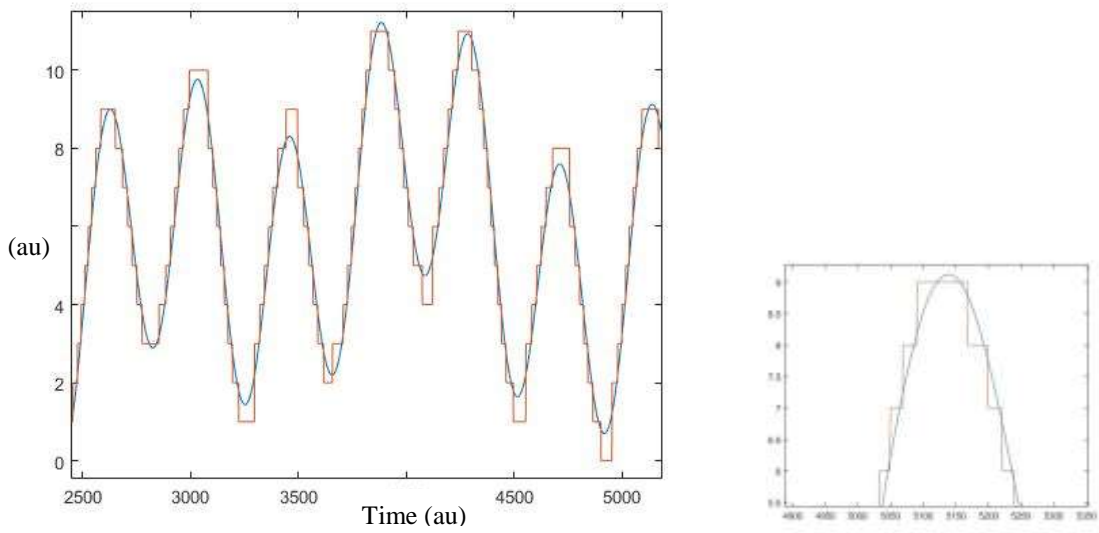

Figure 6. Input (in blue) and output (in red) of twelve-nary system.

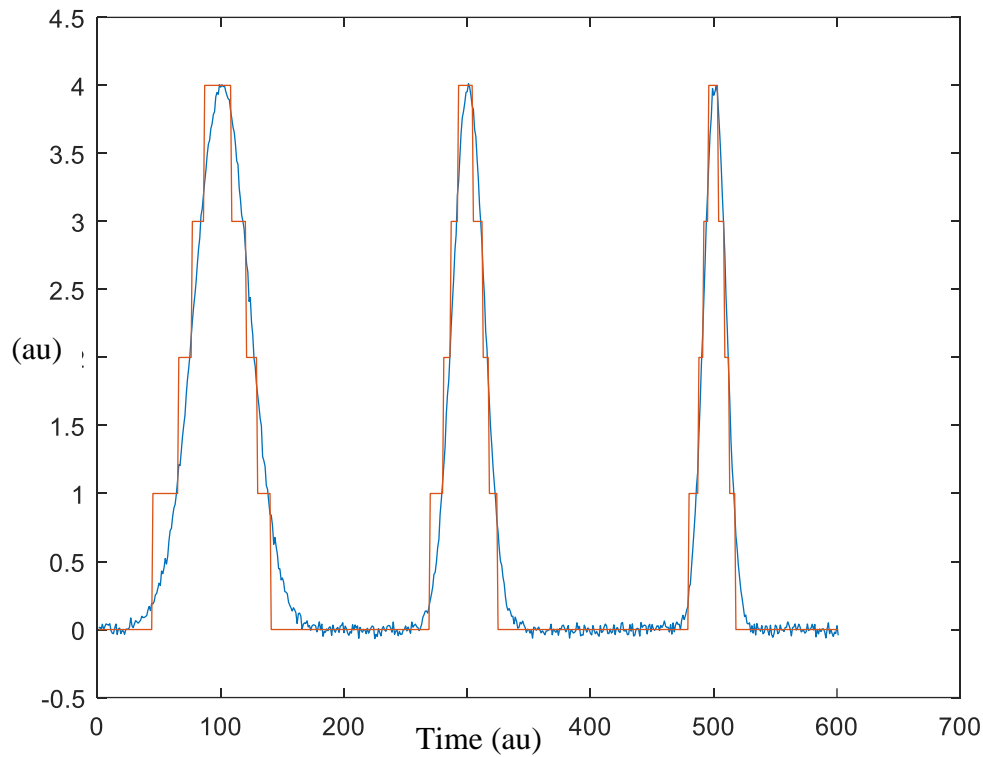

Figure 7.  Three noisy input pulses (in blue) with different rates and their output prosponses (in red).

## *Optical Structure*

An optical structure for the ternary proteretic system is shown in Figure 8.  Each ring laser has three main elements: semiconductor optical amplifier (SOA), optical filter, and optical isolator.  A 3dB directional coupler is used for getting the light in and out of the ring.  Each ring laser is acting as an inverting amplifier.

The input electrical signal modulates the DFB laser via a spatial light modulator (SLM) as shown in Figure 8.  The optical signal carrying the input signal enters the ring lasers 21 and 25.  The inter-connectivity of all rings from 21 through 25 make up the binary proteretic system.  The output of this system enters the second stage via an auxiliary SOA. The second stage has a similar function as the first, completes the ternary system, where the optical output exits from the ring 14 via a 3dB coupler.

This two stage structure can be extended by adding additional stages to implement a multinary proteretic system. A detailed discussion on the binary (single stage) structure can be found in [7].  An all-optical binary proteretic system was implemented and reported in Reference [19].

The optical implementation of the ternary and multinary systems will be considered when a means of funding is available.

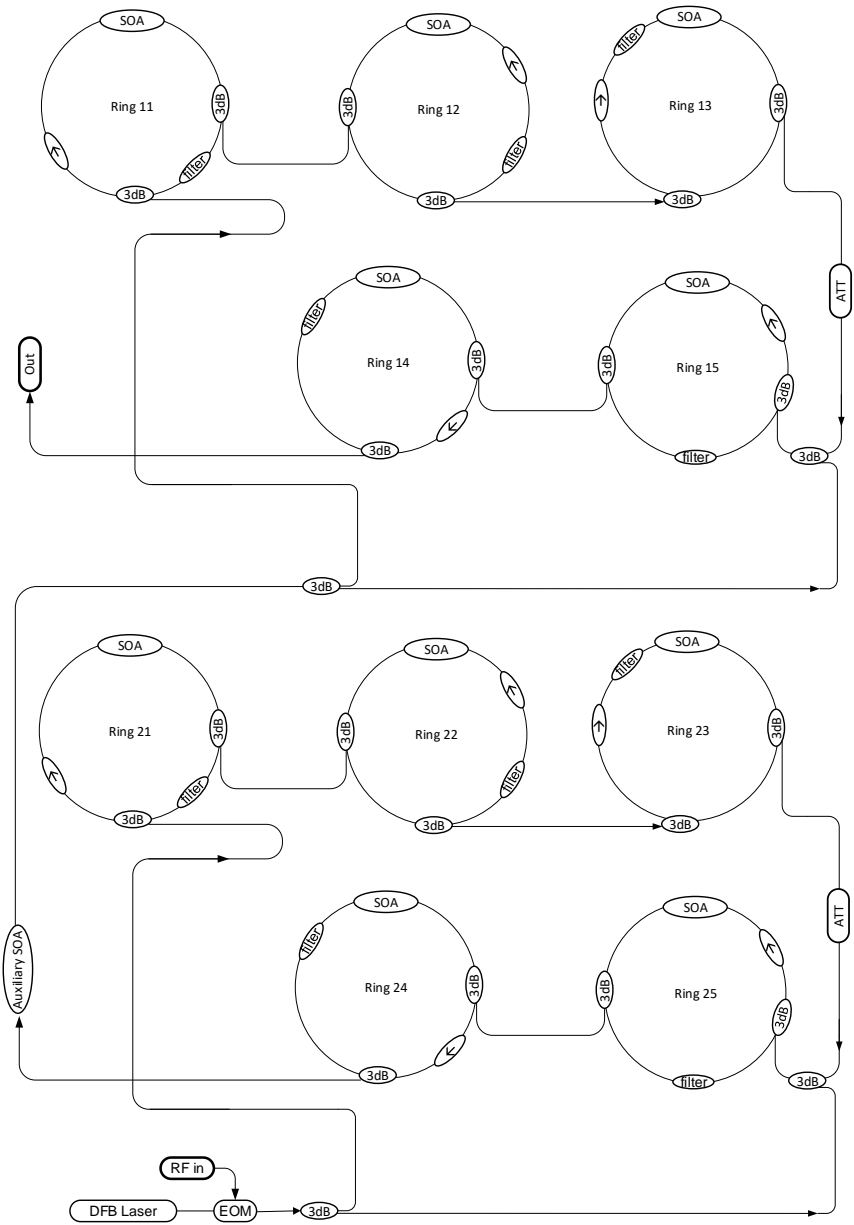

Figure 8. Optical structure for the ternary proteretic system.

## *Philosophical Perspective*

The slight but measurable duration (see Figure 6) by which the output precedes the input could be termed a "habitual overrun," which we call "prosponse."[5] The repetition of an impulse can create a physical situation in which what *might* happen overruns what *does* happen, but does so in a predictive (but not wholly predictable) pattern.[6] The later collapse of the time "bulge" into a thick present (an "event") fulfills a sort of habitual prefiguring of the future.[7] The future snaps back, or slips back (if the convergence is gradual, relative to the proximal timeframes of observers), into a present event that has some durational thickness, however slight. This durational span is not measurable in "clock time," without a kind of abstraction that distorts what is achieved in its endurance, but the span endures such that a form of some kind goes *ahead* of its fulfillment in action.[8] The form is best understood as possibility.[9]

It is easy to see that complex living systems have this proteretic character. For example, if a batter in baseball could not "prospond" to a pitched ball, and begin to act successfully upon the exact place and time (a future "present" with some duration) where a baseball *will be* and might be, in less than a second, it would be impossible to hit the ball "on purpose."[10] Hits would be random, and that is counter to the facts. *Re*sponse cannot occur quickly enough to bring success in complex actions such as hitting a baseball approaching at 160 kph, along a parabolic path, altered by spin and variable resistance of the seams, and conditioned variably by wind, air temperature, and atmospheric pressure, toward a bat being swung in a fractal arc through multiple planes, at variable acceleration[11]; one cannot say that success in such action is some simple, more or less deterministic situation, especially since it is possible to improve one's ability to do it. This is even setting aside the problems of eye-hand coordination under the influence of habit. Human perceptual capacities cannot operate quickly enough for hitters to *perceive* the action they are performing. They must prospond, that is, take a shorter path to the output than our slow vital systems permit. Batters have to bypass the vital system and "get physical," *as* physical as the ball and bat are (and the air, light, wind, temperature, etc.), as material complexes (which to say, electromagnetic complexes). And even at this level of physicality, the possibilities and actualities associated with the flight of the ball can "come apart," between input and output. Something like this happens with the proteretic machine.

Thus far in this example, we are describing the relation between pitcher and hitter in linear terms, and such is also the case with our proteretic electromagnetic wave system, as expressed in light. But as a system (emitter-absorber or pitcher-batter), the relations are nonlinear, existing in a single durational epoch that includes all possibilities and actualities. In order to generalize the event, we treat the whole system as a phenomenon, so we have to think of it as *being observed*, and that complicates matters in well-known ways. We think that the principles of Special Relativity probably do apply to our experiment, in the sense that we must postulate a common timeframe (durational epoch), as a frame of reference, from which to observe the proteretic bulge. Yet, we think this is not different from the presuppositions of the more easily observed hysteretic bulge. Either way, there must be some sort of durational epoch which is treated as "the present" and thereby neutralized for an observer relative to the bulges we then observe.

In the flux,[12] unmanaged by such frames of reference (durational epochs), we think it is reasonable to suppose that, for example, light and sound are still propagated differently. It takes far less energy to propagate light, and far more (indeed, also complex environing conditions, and all their energy as well) to propagate sound waves. But sound waves and light "waves" (let us treat them as such in the flux, even though calling them "waves" requires an abstraction) are undoubtedly part of the natural order, as a temporal order, regardless of their energistic differences (which we interpret more fundamentally as *temporal* differences). Let us not say that light "moves faster" than sound, since we do not want to presuppose such theoretically laden ideas as "velocity" and "inertia," or even "mass," at this point.[13] With energistic differences (or variability, if that is preferred) assumed, we say that high entropy systems (e.g., sound waves) have a *different temporality* than low entropy systems (e.g., the propagation of light). Our task is to interpret that type of temporal difference independent of the energistic differences, if possible. We are here assuming that entropy is an adequate concept to replace the proportional energistic needs or requirements of each system, as it is propagated in time, or more precisely propagated temporally (since time is not a "container" that something can be "in").[14] There may be more to the energistic differences than their differential entropy, but the differences include at least that much in the way of variation.[15] Whether electromagnetic processes can *really* be separated from time (or temporal passage) is a very serious question that we will not attempt to answer. But we are quite confident that temporal and electromagnetic processes can be analyzed and modelled separately to illustrate how they are not identical.

We think that, as a system, the light-time runs *ahead* and the sound-time *behind* the "same event" (if such linear metaphors can be borne) until it reaches a "collapse of the time-function," as we call it.[16] That means that whatever has been abstracted *from* the flux by taking on a differentiated (and hence contrasting) "character" subsides again into the undifferentiated flux, regaining its fully concrete dynamism (or energism, if you prefer).[17] Given any frame of reference at all, therefore, light-time and sound-time (where there are any sonic waves) will collapse into an event, a durational epoch. We can calculate where that collapse occurs for any given observer's perspective, but it is non-local,[18] a cluster of variable occasions that constitute an event, a recovery of the flux as a dynamic form, in the sense that there is no unitary space-time region which is the "where-when" of the collapse of the time function. It can be modeled mereo-topologically. The event is actually somewhere and virtually everywhere,[19] as event, both localized and non-local, which is to say, both unquantized but quantizable in potential and quantized (i.e., actual) in output. Of course, this principle would apply not just to light-time and sound-time, but to the propagation of anything that has an electromagnetic order sufficiently complex to be propagated as a wave (a transference from the flux called an "abstraction").

The collapse of the time function really occurs, we assert, but nowhere and no-when that can be pointed to, except from a single *postulated* perspective, This might this be the proteretic standpoint, and no matter how definite this location appears, as indicated from a single perspective, this "pointing out" is clearly inadequate to "locate" the event. At most the event and the standpoint can be indexed to one another for further comparison and contrast. Such "mapping" can be modeled in binary, but our results suggest multinary (mereotopological) mapping provides a richer understanding of the fact as well as the observed phenomena.

We want to argue that this feature of a balanced system (a feedforward stable system), the nowhere- and no when-ness of the collapse of the time function, may be altered, and hence also detected, in asymmetrical ways in cases of proteresis. From an actual proteretic standpoint (as opposed to a postulated perspective on that standpoint), there is no collapse of the time function. The collapse of time is inseparable from the energy by which the time is made observable. Rather, there is a continuity of possibility that contributes determination, a dynamic form, to the "not yet." The level of determination increases as the proteretic bulge snaps or slides back into the present. Yet, there is contingency until the final threshold, the completion of an event. That issue of finishing an event is determined in the final loss of contingency, and that loss, that egress of contingency, of possibility, must also have some width, some duration, however slight. Here is the proteresis that we believe we see in our machine, see Figure 6. It is not easy to give an interpretation of that "time," and impossible under the standard model of gravitational cosmology. Time has no mass (evidently) and its energy is only an overrun of determination or form, until it collapses.

But such variability in dynamic systems is not limited to just these sorts of factors. Many other actions of living systems (not just human) show proteretic structure. Watching, for example, the battle of the mongoose and a poisonous snake is better described in proteretic terms than in any model of stimulus and response.[20] This proteretic time bulge does not depend on a differential of light and sound in this case; it appears to be entirely an exchange in kinetic energy and light. The mongoose avoids or vacates the place (it is not yet a space) where the snake *will* complete its strike, and hence, defines that space as empty instead of filled (as with the batted ball), *after* the fact, when it has become a "space" in the fulfilled sense. It happens too fast for either animal to "see" it, in any case. We see similar exchanges with frogs that catch insects buzzing through, by uncurling their tongues where and insect will be, fishing cats diving for fish, and so on. It is easy to interpret such examples as an overrun of habitual motion into the future, but to attribute success to chance alone is not reasonable.

Habitual overrun is not just the repetition of actions that have been embedded in the organism as past action. It is not mechanical, even if a mechanism is involved. Rather, habit formation embeds a capacity for prospnse that includes novelty in the action. Possibility is included. The event unfolds according to a proteretic temporal bulge, prosponding to what has not happened yet, and incorporating variation and novelty into the action, right up to the threshold at which the physical systems must collide, or fail to collide.[21] What we have in the proteretic machine is a collision of signals, one from the past and one from the future, defining a space. This fact of living systems, the collisions (not the misses) is difficult to model mathematically, which accounts, perhaps, for its being neglected in physics. But the rate of success, and its corrigibility, nullifies our attributing this sort of physical system to chance alone.

It is not easy to see why living systems are able to prospond unless there is a physical basis for proteresis in simpler physical systems. One is thus faced with two competing hypotheses. Either (1) proteresis is made possible by some macro-level characteristic of the complexity of such (living) systems, or (2) proteresis is a character of all dynamic systems, but the manifestation of proteresis in simpler physical systems is slight and has been difficult to detect and impossible to measure, and has thus been dismissed, neglected, or treated as part of the noise in the system. It became possible to observe and perhaps measure the phenomenon (in simpler systems) only in the last decade or so. We have managed to create a proteretic bulge in a very simple system. The issue is what does it mean, and how is the bulge to be measured? How is it possible to devise a unit of measure for proteresis?

We take the provisional view that proteresis is a characteristic of *all* dynamic physical systems. There is more than an analogy between the proteresis of living systems and the proteresis shown by the proteretic machine. Examples from living systems help us understand the phenomenon in non-living systems. Even though this is more than an analogy, we think, it may be stated as one, heuristically. Assuming that we do not treat spacetime as a primitive of the physical universe (and there is good reason to set this idea aside[22]), then we must have an idea that does similar work. The concept of a minimal spacetime region may have some usefulness in modeling gravitational forces, but here we assume the "independence of time." Electromagnetic fields have no rest-mass, as such (only associated mass), and are hence reversible (to great energistic effect, as Tesla showed).[23] Our model adopts such a viewpoint. The closest reason we can offer for the masslessness of electromagnetic fields is that they do not exist at rest, and this idea may not be the flux itself, but it is surely a close abstraction. We have a highly "informative" 0 (maximal information) here. We know that living systems cannot repeat their past actions with much exactitude; yet, why are we surprised when simpler systems also don't repeat themselves exactly? Why would we expect a simple (non-living) dynamic system to be able to repeat (approaching perfectly and without remainder) its earlier "state." Is that a reasonable expectation? A term like "state" is a generalization across real temporal changes, postulated for whatever purposes may be inherent in some inquiry. But no physical system, no matter how simple, precisely repeats an earlier pattern or mode. It is fair to speak of a "state of a system" only conditionally, then. In fact, everything is changing all the time.

With simple physical systems, we have more reason to expect them to be capable of both hysteresis and proteresis than to expect only hysteresis. If hysteresis alone characterizes them, we cannot explain novelty in physical systems.[24] The undeniable reality of novelty provides a serious motive to find a form of physical exchange that is best modelled and explained proteretically. What we seek is not just an electronic or digital mimicry of proteresis. Such models have been created already. They show how we can mimic proteresis in digital environments because either digital environments share some essential and irreducible (if complex) overlap with living systems, or because such systems take on proteretic character at some threshold of complexity. This latter has been the favored view, and is somewhat optimistically called "artificial intelligence." The more reasonable view is that both living and digital dynamic systems share with simpler physical systems a proteretic capacity to overrun the present and act upon the future.

*Discussion*

In the plainest language, the proteretic machine as an anticipatory structure shows the *form* of the future, fully actual, occurring before the *content* arrives. The first inverting agent (device) represents the "inverted past" (essentially a model of future form projected from the past) which is added (mixed or diffused) with the present (input); and then the result is "moved to the past" (or "released," as it were, like exhaust) by the second agent. How "far" a system might overrun the present (going into the future) depends on how deeply the first inverting agent can go back into the past. But by "depth" we do not mean "length of time." Let us clarify.

Here we find a surprising result. One would think that going further into the future requires going further into the past, in terms of temporal spans, so the "more past" will yield "more future." But the system works the other way. Going further into the future for the first agent makes the second agent go less and less temporal distance into the past; the densest past, in terms of the information prefigured in proteresis is the immediate past, not the more distant. This means that the total "lag time" of the first agent and the second agent is conserved according to some ratio involving relevance, density of information, and a threshold. The durational range and its limits in the proteretic machine may have to do with the speed of light in a given electromagnetic environment, which, as we know, is variable in its waveform across the light spectrum. This limitation is due to its being an optical device, with physical constraints. But other such ratios may accompany other types of physical systems. (One thinks of the Doppler effect in waves.) Perhaps, then, infrared waves provide a different durational span than ultraviolet waves, on this interpretation. The differences would be small but should be measurably different. To get "more future" thus requires "more" *inverted* past, a deeper dive into the immediately prior event, its trove of information, not "more" past time as such. It is inverting the past (signal or impulse) that opens the future to a field of possibility.

So, if we are right, the trick to understanding the structure of possibility, in this case, is the question: what is the inverted past? Perhaps it is the events that did not get actualized.[25] "1" is actual, and "0" is possible. But "1" as "event," is more than 1, it is 1 + 0, and the denser the 0, the more telling is the "1." A future action consists of *mixing* the present with the *inverted* past and then *going back* more or less "deeply" (let us call it "intensity") into the immediate past, and bringing the full "load" into the dynamic form that has already become actual, but is not yet actual *for* the event unfolding, enduring, as it were. That "load" would inevitably include some of the possibilities that *did not* get actualized in the immediate past. So, some 0 is in the fulfilled 1 that will occur when the content "catches up" to the form (the collapse of the time function). That means that in the present duration 1 ≠ 1, but rather seems like it soon will *just be 1*, an identity. This "not yet" of the present would have to be modeled as an open "set."[26] It seems to follow that the inverted past makes the present like an emulsion (saturated mixture) of the inevitable and the possible (of 1 and 0).[27] These present possibilities would be independent of future possibilities, since they already *failed* to happen, but some of the information could be projected into the future and overlap with what *might still* happen. The unactualized possibilities, some portion of the "load" mentioned above, remain possible. If the overlap is sufficient, we could use the (properly selected) 0 in the present to predict or otherwise describe the future.

Under our interpretation of proteresis, the relation of past-present-future is quantized in only one direction, which is itself not wholly fulfilled or satisfied, due to the conserved 0 with which it is mixed to the threshold of thermodynamic "emulsion."[28] In short, nothing is "over" or "finished." We thus avoid Heisenberg's form of uncertainty for a different kind of uncertainty. And with the presence of such inverted past mixed with the enduring, unfinished event, we have delayed indefinitely the full collapse of the wave function because, although the light emitted by the proteretic machine is quantized, it is also "unfinished," being fully "1" but enhanced by unused information, some of which is still usable. It is present in a usable quantum, but includes a dynamism that is already in constructive relation with overlapping possibility "clusters" in the near future.[29] One could call this relation "lure" based on "sympathy" (the overlap).[30] The future, considered as actively approaching the present, has a possibility structure that exerts a "lure" (not a force) upon the present.

This idea of one-dimensional quantization can be extended to multi-dimensional electromagnetic situations as the vector quantization. There can be "trends."[31] Then the output of the inverted device can be opened to many clusters of possibilities in the near future. These can go beyond the simple physical system as "information," and would explain the continuity of physical systems with physiological systems. What is wanted is a model of how past possibilities accompany the actual past into the present when the past is inverted.

In regard to the physiological aspects of such time-systems, which we may treat as very high entropy systems, neurons possess hysteretic behavior, see for example Mayergoyz, C. Korman 2020 [9]. The proteretic characteristic of neurons should also be capable of demonstration; and neuronal behavior is possibly the reason for the fast perception/thought processes in spite of slow behaving neurons (see M. R. Sayeh and I. Calmese 2008 [8]). If we can show this proteretic feature of neurons, that capacity may also account for why it is possible for animals (including humans) to interact with the environment by slightly "seeing" (in the sense of perceiving and/or sensing) the future.[32] We have known *that* we do this "seeing" of the future (one must in order to hit a baseball), but until now, we have not known what to look for in neuronal interactions. Now we do, and it is continuous with simpler physical systems. Clearly it is possible to teach a human being (and perhaps eventually a machine) to hit a fastball, and that kind of learning shows a manipulation of proteretic functions.

Bringing this proteretic fact and phenomenon to the market of applications, the binary version of the proteretic device has been simulated in electronics and in photonics platforms [10,7]. Recently the all-optical proteretic device has been implemented with photonic discrete components (SOA based) [18,19]. Exploiting this new technology, faster computers and more efficient cellphones among other things will be built. There are clear applications for artificial intelligence. If the inverted past is the key, and the future "opens" to the device in proportion to whatever the electromagnetic situation affords, then we must learn about the inversions that come together in physiological systems and reproduce them in non-physiological systems, which is to say, lower entropy systems. That may be possible now, since we know what to look for in physiological (high entropy) systems.

---

[1] We would apologize for using a baseball illustration (recognizing that not everyone appreciates baseball), but a baseball provides a convenient "middle-sized system" that travels and is acted on at the limit of human perception. This makes it a tempting example. We note that Richard Feynman was fond of baseball examples for similar reasons (see 1985, p. 87, for example). He is far from being the first physicist to notice how natural are the analogies. We aim to push beyond analogy, however.

[2] We are using the verb "to experience" for lack of a better one that will cover both what happens when a signal reaches its output, e.g., when a light beam is absorbed by a detector, and a living system is stimulated by some sort of energy.

[3] Robert Rosen repeated (evidently without knowing it) a research program that had been largely accomplished half a century earlier by A.N. Whitehead. See Herstein, 2006, pp. . . .

[4] The delta symbol here must be thought of in a slightly different way from its normal interpretation as time $t_1$ to $t_n$. In this case delta should be read as time $t_n$ to time $t_1$. This re-interpretation of delta will be taken up later in the paper.

[5] The choice of "habit" is motivated by the description of C.S. Peirce, who says that the universe itself is a "habit." See the analysis of John F. Miller III, in "The Role of Habits in Peirce's Metaphysics," in *Southwestern Journal of Philosophy*, 9:3 (Spring 1978), 77-85. He says: "Through this one concept [habit], Peirce's analyses of belief, the categories, laws of nature, evolution, meaning, thinking, the doctrine of continuity (synechism), fallibilism, tychism

[the doctrine of chance] are unified into a systematic, pragmatic conceptual framework." (77) William James, John Dewey, and many other subsequent inquirers made philosophical use of Peirce's idea of the universe as a habit. Peirce did not intend the term metaphorically, and we do not either, but a fuller explanation must await a later inquiry.

[6] We take this assertion to be in keeping with the ideas being suggested by David Deutsch and those working at the Clarendon Laboratory at Oxford, on constructor theory. In a subsequent paper, we will attempt to join quantum information theory with the theory of electromagnetism. For now we discover that we need the idea of "matter," but in a sense that is non-classical. Henri Bergson defines "matter" as an aggregate of images that repeat with almost no variation. His sense of the word "image" is unfamiliar to most contemporary inquirers, but the point is that *repetition* of an event with an absolute minimum of variation is a fair way to think about matter without abstracting the concept overly and drawing it away from the temporal flux. See Bergson, 1988, 9-16. For a fuller account of the meaning of "possibility" as used here, see Auxier and Herstein (2017), chapters 7-9.

[7] For an account of this sense of "event," see Auxier and Herstein 2017, Introduction; Whitehead, 1978, p. 73.

[8] A.N. Whitehead says, "a duration is a slab of time with temporal thickness [that] is the final fact of observation from which moments and configurations are deduced as a limit which is a logical ideal of the exact precision inherent in nature." (1922, p. 7)

[9] Again, we see this as in keeping with David Deutsch's (and Clarendon's) work on possibility. See below.

[10] It is important to keep in mind that the flight of such a ball can be interrupted, however unlikely that may be. Here is a video on an instance when it happened: https://www.youtube.com/watch?v=1PyCpG06138. A bird swoops between the pitcher and batter at the precise moment the ball occupies the same durational bulge. The result is the collapse of the time function, as we will explain. Please note that even though the ball never "arrived," the catcher "caught it" anyway. We would say he caught the possibility, but not the ball. It appears that the batter in this instance had decided not to swing, since the interruption in the ball's flight comes late enough that it would have been impossible for it to be hit by the batter. But it is clearly possible the batter could have swung. In fact, catchers in baseball catch "possibilities" all the time, when the ball is actually hit. One will see them close the mitt on the possible ball every time it is hit. There is a concrete relation between the possible and the actual that *can* come apart. Feynman says as much when he points out that unless a detector is perfect, there are really three possibilities in the classic double slit experiments, the third being that the particle was not absorbed at all. See Feynman 1985, p. 82n. A.N. Whitehead calls these relations between what is actual and the associated possibilities "fundamental feelings" and says their character is that of a vector (1978, p. 55). We "feel" possibilities, and he does not mean this to be limited to living systems. It is a character of all physical systems. See also Altschul and Biser, 1948, pp. 12-13. Feynman points out that this kind of analysis does an end run around the uncertainty problem, 1985, pp. 55-56n. That is how we are thinking of these relations here. We realize that the term "feeling" is historically loaded with teleological meanings that are not relevant to this investigation. We will not draw on this aspect of their meaning at all. The "feelings" merely illustrate complex, dynamic physical systems for us. We also use the term "purpose," but again, this has to do with a characteristic of strictly physical systems, and is based on Whitehead's use of the term "physical purpose," which is non-teleological and mathematical. See Whitehead 1978, pp. 276-277. We avoid using the notion of a space-time region from the Standard Model of Gravitational Cosmology because that notion is seriously in doubt at present. We use "the durational bulge" instead. The trouble with "spacetime" is widely discussed. See the public lectures by, for example, Nima Arkani-Hamed, "The End of Spacetime," https://www.youtube.com/watch?v=t-C5RubqtRA, accessed May 30, 2020; and Chiara Marletto, "Why Classical Ideas of Gravity Are Wrong," https://www.youtube.com/watch?v=t-C5RubqtRA. We notice that Marletto's ideal of extending her non-classical interpretation of super-position to something that weighs 2 kg. (Schroedinger's cat) is not so very different from our baseball illustration. Only recently has it become possible to question the dogma of general relativity and its associated jungle of concepts, but finally it is happening on a wide scale. See also from Clarendon, J. Fullwood and V. Vedral (2025) [32].

[11] See R. Cross (2009).

[12] See Alfred North Whitehead, *The Principle of Relativity* (1922), p. 10 for the meaning of "flux."

[13] See an example of how one might model "mass generation" in keeping with the standard model, by Gunther Kletetschka, "Three-dimensional Time: A Mathematical Framework for Fundamental Physics" (2025) [33], https://www.worldscientific.com/doi/epdf/10.1142/S2424942425500045, accessed July 3, 2025.

[14] See Charles M. Sherover, "Are We *in* Time?" 2003.

[15] If time and energy can be analyzed separately, we don't need Dirac's "anti-electrons," proposed in 1931, but we would clearly understand his motive for proposing them, as with tacheons and anti-matter, and so many other ideas that would grant the existence of some sort or reciprocal or negative complement to time's arrow. We are interpreting positive possibility as the complement.

[16] The collapse of the time function is what we will analyze and model mathematically in the follow-up paper. To do so here takes us too far away from the present purpose of interpreting the "time machine."

[17] We realize this language is unfamiliar within the standard model of gravitational cosmology. Our usage is taken from Whitehead who says: "the flux of time is essential to the concrete reality of nature, so that a loss on the time-flux means transference to a higher abstraction." (1922, p. 10) By "abstraction" he does not mean the mental abstraction of an observer, but rather the rarification of the flux itself into differentiated orders of time, such as we find in the character of sound as contrasted with the character of light. Both are "abstractions" of the flux. In his formalization, "the electromagnetic equations adopted are Maxwell's equations modified by the gravitational tensor components in the well-known way. Light is given no privileged position, and all deductions concerning light follow directly from treating it as consisting of short waves of electromagnetic disturbance." (1922, p. 10) This framework explains our findings much more closely that the standard model of gravitation. Even if light deserves a privileged position, in the final analysis, we think there is better reason to treat it as *included in* the order of nature than as an ideal limit *to* nature. The standard model would think our results impossible or illusory. That is a limitation of that model, not of our results. If all time is illusory, then let us study the illusion.

[18] Altschul and Biser, in discussing Born's statistical model of the "collapse," say "to obtain the exact probability of locating the electron [or any other wave function, we generalize] within a tiny volume dxdydz, $\psi_n \psi_n$*dxdydz is integrated throughout the space yielding: $\iiint \psi_n \psi_n$*dxdydz = 1 which expresses the certainty that the electron [or any other wave function] will be found at all points in some volume. (The triple integral equated to one is precisely the normalization condition in Schroedinger's theory)." (p. 18) We must set aside the question of how to generalize this non-local presence within a tiny volume, recognizing that the term "volume" requires an interpretation. This issue will be taken up in a follow-up paper.

[19] This is Whitehead's language. See Auxier and Herstein, p. 135; Whitehead 1978, p. 284.

[20] It is true that the mongoose is immune to the poison of even the deadliest snakes, such as the black mamba it battles here, but that does not change the character of the fight, which clearly involves prospone, and interpreted systemically, it depends on proteresis. https://www.youtube.com/watch?v=yRowC6t8tjA. Altschul and Biser argue that differences we measure in the social sciences are directly "given" (i.e., observable), and thus unique actions can be generalized with their differences set aside more easily than unique events in quantum physics, which must be interpreted before they can be generalized. (See Altschul and Biser, 1948) We agree.

[21] This threshold requires further study and it will become the main theme of a subsequent study. We must perform further experiments with the time machine to understand it better.

[22] Numerous studies have worked around treating spacetime as a primitive, but see R. Kastner, "The Relativistic Transactional Interpretation and Spacetime Emergence," [34] https://arxiv.org/abs/2103.11245, accessed July 3, 2025.

[23] Starting as simply as possible, perhaps this unit can be sought as a B-field, a magnetic flux density, so as to satisfy the Lorentz transformations.

[24] Although our thinking follows the DeBroglie-Bohm theory, it is not deterministic, due to the inclusion of possibility in interpreting the wave function.

[25] A more thorough discussion of possibility in this sense is in Auxier and Herstein, 2017, chs. 7-9. See also M. Weber's summary, 2017.

[26] In light of other work recently done, it would be better to call this an open class. See Auxier, 2020, chs. 20-24.

[27] One may conjecture a mixture characterized by Brownian movement. See A. Einstein, 1906, §1-2. The advantage of Einstein's treatment is that it generalizes above fluids and gases and takes the movement as a thermodynamic system. Altschul and Biser make a similar point (1948, p. 17).

[28] See Einstein, 1958 [1908], pp. 81-82.

[29] Here we think our idea approaches and perhaps blends with and adds to Deutsch's constructor theory.

[30] The term "lure" is again borrowed from Whitehead, 1929, see p. 88.

[31] These trends would not rule out highly improbable future events, which have come to be called popularly "Black Swans," following the usage of N. Taleb.

[32] Again, see Bergson's modifications of Fechner's Law, *supra*.